# Classical emulation of a quantum computer

Brian R. La Cour*, Corey I. Ostrove, Granville E. Ott,
Michael J. Starkey and Gary R. Wilson

*Applied Research Laboratories,*
*The University of Texas at Austin,*
*P. O. Box 8029 Austin, Texas 78713-8029, USA*
**blacour@arlut.utexas.edu*

This paper describes a novel approach to emulate a universal quantum computer with a wholly classical system, one that uses a signal of bounded duration and amplitude to represent an arbitrary quantum state. The signal may be of any modality (e.g. acoustic, electromagnetic, etc.) but this paper will focus on electronic signals. Individual qubits are represented by in-phase and quadrature sinusoidal signals, while unitary gate operations are performed using simple analog electronic circuit devices. In this manner, the Hilbert space structure of a multi-qubit quantum state, as well as a universal set of gate operations, may be fully emulated classically. Results from a programmable prototype system are presented and discussed.

*Keywords*: Quantum computing; analog electronics; signal processing.

## 1. Introduction

In 1980, Paul Benioff first introduced the concept of a quantum computer by demonstrating that a closed quantum system could be used to model the general process of computation.[1] The idea was later formalized by David Deutsch through the concept of a universal quantum computer as a generalization of a classical Turing machine.[2] Nearly a decade later, Peter Shor demonstrated that quantum computers could be used to factor large numbers more efficiently than any known classical algorithm and such devices could be designed with fault tolerance.[3,4] Quantum computers thereby offer a challenge to the strong Church–Turing thesis that one can do no better than a classical Turing machine.[5,6] This paper examines the nature of this presumed "quantum speedup" and whether, by broadening our notion of a classical computer, one may in fact emulate and exploit it.

*Corresponding author.

Current approaches to quantum computing use, of course, true quantum systems, such as photons, trapped ions, or superconducting circuits.[7,8] All of these approaches rely upon maintaining a highly coherent quantum state through a series of gate operations in order to achieve a computational advantage. Preparing and manipulating such systems can be quite difficult, as small interactions with the environment quickly lead to decoherence of the state and, consequently, a significant loss in performance.[9] The recent discovery of classical analogues to quantum systems has suggested that a classical emulation of a quantum computer may be feasible and both easier to build and far less susceptible to decoherence.[10–13]

Motivated by this insight, we have developed a novel approach to quantum computing using classical analog signal processing. This approach uses a signal model that is mathematically equivalent to a multi-qubit, gate-based quantum computer.[14] Within this model, we have devised a novel scheme for addressing individual qubits, or groups of qubits and applying gate operations upon them using analog electronic adders, multipliers and filters. We furthermore have constructed a model of quantum measurement gates based on amplitude threshold detections that is capable of reproducing phenomena, such as quantum contextuality and entanglement, thought to be uniquely quantum in nature and important to quantum computing.[15,16] Thus, using this approach it is possible to realize any particular quantum computing algorithm or protocol, including Shor's factoring algorithm, Grover's search algorithm, fault-tolerant quantum error correction and even quantum teleportation. Finally, an actual hardware demonstration system has been constructed that is capable of emulating a two-qubit quantum device.

The organization of the paper is as follows: In Sec. 2, we describe the basic physical representation for quantum emulation and then turn to a hardware implementation of a two-qubit device in Sec. 3. The fidelity of this device is analyzed in Sec. 4, and the prospects for a larger-scale device are discussed in Sec. 5. Finally, Sec. 6 summarizes our findings and conclusions.

## 2. Physical Representation

This section describes the notional physical representation of quantum states and various gate operations. These are described in more detail elsewhere.[14] Here, we summarize the description for completeness and, later, turn to a specific hardware implementation. We begin with a description of the mathematical Hilbert space of an $n$-qubit quantum system, then relate these mathematical constructs to a classical, signal-based representation.

### 2.1. *Hilbert space description*

The state of an $n$-qubit quantum computer may be represented by an element $|\psi\rangle$ of a $2^n$-dimensional Hilbert space $\mathcal{H}$ taking on the particular form of a tensor product of $n$ two-dimensional Hilbert spaces $\mathcal{H}_0, \ldots, \mathcal{H}_{n-1}$ such that $\mathcal{H} = \mathcal{H}_{n-1} \otimes \cdots \otimes \mathcal{H}_0$, where $\otimes$ is the tensor product. A single element of one of the $n$ constituent Hilbert spaces constitutes a qubit. The specification of an inner product $\langle\phi|\psi\rangle$ between states $|\phi\rangle$ and $|\psi\rangle$ in $\mathcal{H}$ completes the Hilbert space description.

We shall denote by $|0\rangle_i$ and $|1\rangle_i$, a pair of orthonormal basis states, termed the computational basis, for $\mathcal{H}_i$ and $i \in \{0, \dots, n-1\}$. Taking tensor products of these individual basis states, we obtain a set of $2^n$ orthonormal basis states for the product space, $\mathcal{H}$. A particular binary sequence $x_0, \dots, x_{n-1}$ therefore corresponds to a single basis state $|x_{n-1}\rangle_{n-1} \otimes \cdots \otimes |x_0\rangle_0$. For brevity, this binary sequence may be represented by its decimal form, $x = x_0 2^0 + \cdots + x_{n-1}2^{n-1} \in \{0, \dots, 2^n - 1\}$, so that the corresponding basis state may be written succinctly as $|x\rangle$ or, more explicitly, as $|x_{n-1}, \dots, x_0\rangle$. Let $\langle x|\psi\rangle = \alpha_x \in \mathbb{C}$ for a given state $|\psi\rangle \in \mathcal{H}$ and basis state $|x\rangle$. This state may then be written

$$|\psi\rangle = \sum_{x=0}^{2^n-1} \alpha_x |x\rangle. \tag{1}$$

## 2.2. *Signal-based representation*

A key concept in our classical representation of a quantum state is the notion of *in-phase* and *quadrature* signals, which may be used to encode two distinct signals within one signal. Suppose we have a complex number $\alpha = a + jb$ with $a, b \in \mathbb{R}$. Physically, the real and imaginary values may be represented by, say, a pair of distinct direct current (DC) voltages. It is also possible to encode them in a single alternating current (AC) voltage signal $s(t)$ of carrier frequency $\omega_c > 0$ such that

$$s(t) = \mathrm{Re}[\alpha e^{j\omega_c t}] = a\cos(\omega_c t) - b\sin(\omega_c t). \tag{2}$$

Given $s(t)$, the parameters $a$ and $b$ can be recovered by the following procedure. First, we split or "clone" the signal into two copies, which can be done classically, and then multiply each by $2\cos(\omega_c t)$ and $-2\sin(\omega_c t)$, respectively, to obtain

$$2\cos(\omega_c t)s(t) = a + [a\cos(2\omega_c t) - b\sin(2\omega_c t)], \tag{3a}$$

$$-2\sin(\omega_c t)s(t) = b - [b\cos(2\omega_c t) + a\sin(2\omega_c t)]. \tag{3b}$$

Multiplication by the quadrature components thus creates signals at the sum and difference frequencies $2\omega_c$ and 0 (i.e. DC). Low-pass filtering these two signals, then, yields the coefficients $a$ and $b$, as desired.

A similar approach may be used to encode two complex numbers $\alpha$ and $\beta$, and hence a single qubit, using the complex quadrature signals $e^{j\omega_0 t}$ and $e^{-j\omega_0 t}$. Physically such a choice would, again, correspond to using two distinct real signals, each representing the real and imaginary parts, to realize the complex signal. With this "dual rail" representation, the corresponding complex signal is given by

$$\psi(t) = \alpha e^{j\omega_0 t} + \beta e^{-j\omega_0 t}. \tag{4}$$

As before, the coefficients $\alpha$ and $\beta$ may be recovered (as pairs of DC voltages) by multiplying copies of $\psi(t)$ by $e^{-j\omega_0 t}$ and $e^{j\omega_0 t}$, respectively, and then low-pass filtering. Note that multiplication, in this case, is *complex* multiplication between two dual-rail signals, which results in a dual-rail output.

More generally, we may identify the single-qubit basis states $|0\rangle_i$ and $|1\rangle_i$, for qubit $i$, with the basis functions $\phi_0^{\omega_i}$ and $\phi_1^{\omega_i}$, where $\phi_0^{\omega_i}(t) = e^{j\omega_i t}$ and $\phi_1^{\omega_i}(t) = e^{-j\omega_i t}$ are the in-phase and quadrature signals, respectively. For $n$ qubits, the basis state $|x\rangle$ is represented by the basis signal $\phi_x$ composed of a product of $n$ single-qubit signals as follows:

$$\phi_x(t) = \phi_{x_{n-1}}^{\omega_{n-1}}(t) \cdots \phi_{x_1}^{\omega_1}(t) \cdot \phi_{x_0}^{\omega_0}(t), \tag{5}$$

where $\phi_0^{\omega_i}(t)$ and $\phi_1^{\omega_i}(t)$ are defined above. Thus, function multiplication serves as a tensor product between qubits. Unlike the Kronecker product of matrices, though, the order is unimportant, as the qubits are distinguished by their distinct frequencies. Note that the spectrum of $\phi_x$ will consist of the $2^n$ sums and differences of the $n$ component frequencies, which represent the Hilbert space. We refer to this description as the quadrature modulated tonals (QMT) representation.

By way of convention, we take $0 < \omega_0 < \cdots < \omega_{n-1}$, where $\omega_i = 2^i \omega_0$, and refer to this as the *octave spacing scheme*. The quantum state $|\psi\rangle$ can now be represented as a complex, $n$-qubit signal $\psi$ which, at time $t$, is given by

$$\psi(t) = \sum_{x=0}^{2^n-1} \alpha_x \phi_x(t). \tag{6}$$

For two such signals $\phi$ and $\psi$, the inner product is defined to be

$$\langle \phi | \psi \rangle = \frac{1}{T} \int_0^T \phi(t)^* \psi(t) dt, \tag{7}$$

where $T$ is a multiple of the period $2\pi/\omega_0$ of the signal. Note that the inner product corresponds to a low-pass filter, and $\langle \phi_x | \psi \rangle = \alpha_x$ represents a pair of DC values giving the components of the quantum state for the $|x\rangle$ basis state. This completes the Hilbert space description, thereby demonstrating the mathematical equivalence of this representation to that of a true multi-qubit quantum system.

### 2.3. *Gate operations*

In our approach, subspace projections are used for performing gate operations. Given a quantum state $|\psi\rangle \in \mathcal{H}$, we can mathematically decompose it into the two orthogonal subspaces corresponding to, say, qubit $i$ as follows:

$$|\psi\rangle = \Pi_0^{(i)} |\psi\rangle + \Pi_1^{(i)} |\psi\rangle = |0\rangle_i \otimes |\psi_0^{(i)}\rangle + |1\rangle_i \otimes |\psi_1^{(i)}\rangle, \tag{8}$$

where $|\psi_0^{(i)}\rangle$ and $|\psi_1^{(i)}\rangle$ are the $(n-1)$-qubit *partial projection* states.

A linear gate operation on a single qubit may be represented by a complex $2 \times 2$ matrix $U$, where

$$U = \begin{pmatrix} U_{0,0} & U_{0,1} \\ U_{1,0} & U_{1,1} \end{pmatrix}. \tag{9}$$

If $U$ acts on qubit $i$ of state $|\psi\rangle$, then the transformed state is

$$|\psi'\rangle = [U_{0,0}|0\rangle_i + U_{1,0}|1\rangle_i] \otimes |\psi_0^{(i)}\rangle + [U_{0,1}|0\rangle_i + U_{1,1}|1\rangle_i] \otimes |\psi_1^{(i)}\rangle. \tag{10}$$

Thus, the gate operation is applied only to the addressed qubit basis states, not to the partial projections. This, of course, is only a mathematical operation. A physical method of construction is needed to realize the transformation.

In our QMT representation, a pair of complex signals $\psi_0^{(i)}(t)$ and $\psi_1^{(i)}(t)$ corresponding to the partial projection states $|\psi_0^{(i)}\rangle$ and $|\psi_1^{(i)}\rangle$ is produced by taking the initial complex signal $\psi(t)$, multiplying copies of it by $\phi_0^{\omega_i}(t)$ and $\phi_1^{\omega_i}(t)$, respectively, and passing them through a pair of specialized band-pass filters that output the desired projection signals.[14] Given this pair of complex signals, along with the complex, single-qubit basis signals $\phi_0^{\omega_i}(t)$ and $\phi_1^{\omega_i}(t)$, we may construct the transformed signal $\psi'(t)$ using analog multiplication and addition operations as follows:

$$\psi'(t) = [U_{0,0}\phi_0^{\omega_i}(t) + U_{1,0}\phi_1^{\omega_i}(t)]\psi_0^{(i)}(t) + [U_{0,1}\phi_0^{\omega_i}(t) + U_{1,1}\phi_1^{\omega_i}(t)]\psi_1^{(i)}(t). \tag{11}$$

Two-qubit gate operations, such as Controlled NOT (CNOT) gates, may be constructed similarly.

Importantly, this approach to performing gate operations requires only a single subspace decomposition of the original signal into two constituent signals and does not require a full spectral decomposition, as would be required if one were performing an explicit matrix multiplication operation over the entire $2^n$-component state. This approach provides a significant practical advantage to implementation and more closely emulates the intrinsic parallelism of a true quantum system.

### 2.4. *Measurement gates*

The procedure for performing measurements is quite similar to that for performing gate operations. To perform a measurement on, say, qubit $i$, we construct the partial projection signals $\psi_0^{(i)}(t)$ and $\psi_1^{(i)}(t)$, as before, and measure their root-mean-square (RMS) values, given by

$$q_0^{(i)} = \frac{1}{T}\int_0^T |\psi_0^{(i)}(t)|^2 dt, \quad q_1^{(i)} = \frac{1}{T}\int_0^T |\psi_1^{(i)}(t)|^2 dt. \tag{12}$$

This can be done more easily by adding the real and imaginary parts of, say, $\psi_0^{(i)}(t)$, measuring the RMS value of the sum, and then squaring the result, since

$$\frac{1}{T}\int_0^T [\mathrm{Re}\psi_0^{(i)}(t) + \mathrm{Im}\psi_0^{(i)}(t)]^2 dt = q_0^{(i)}. \tag{13}$$

According to the generalized Born rule, the outcomes 0 and 1 occur with probability $p_0^{(i)} \propto q_0^{(i)}$ and $p_1^{(i)} \propto q_1^{(i)}$, and these probabilities may be computed explicitly through analog sum and division operations. For each such qubit measurement, a random input DC voltage representing a random number $u_i$, chosen uniformly in the interval $[0, 1]$, may be input to a comparator device such that when $u_i > p_0^{(i)}$, a binary outcome of 1 is obtained with a probability given by the Born rule.

To measure a second qubit, the same procedure is followed but using the (unnormalized) "collapsed" state $\Pi_0^{(i)}|\psi\rangle$ or $\Pi_1^{(i)}|\psi\rangle$, depending upon whether outcome 0 or 1, respectively, was obtained in the first measurement. The selection of the collapsed state may be implemented through a simple switch controlled by the binary measurement output. This procedure may be repeated until all $n$ qubits are measured. Doing so results in an $n$-bit digital output whose distribution follows the quantum mechanical predictions, at least to the limits of hardware fidelity.

## 3. Hardware Implementation

We have implemented in hardware a device capable of initializing the system into an arbitrary two-qubit state and operating one of a universal set of gate operations. A picture of the current hardware setup is shown in Fig. 1. We use a signal generator to produce a baseline 1000 Hz tonal, from which all other signals are generated and thereby phase coherent. The lower frequency qubit, labeled $B$, is taken from the signal generator, with a separate, 90° phase-shifted signal used to represent the imaginary component. The higher frequency qubit, labeled $A$, is derived from qubit-$B$ via complex multiplication, which results in frequency doubling. Thus, $\omega_A = 2\pi(2000\,\text{Hz})$ and $\omega_B = 2\pi(1000\,\text{Hz})$. The two single-qubit signals are multiplied to produce the four basis signals $\phi_{00}(t) = e^{j(\omega_A+\omega_B)t}$, $\phi_{01}(t) = e^{j(\omega_A-\omega_B)t}$, $\phi_{10}(t) = e^{j(-\omega_A+\omega_B)t}$, and $\phi_{11}(t) = e^{j(-\omega_A-\omega_B)t}$ centered at frequencies +3000 Hz, +1000 Hz, −1000 Hz, and −3000 Hz, respectively.

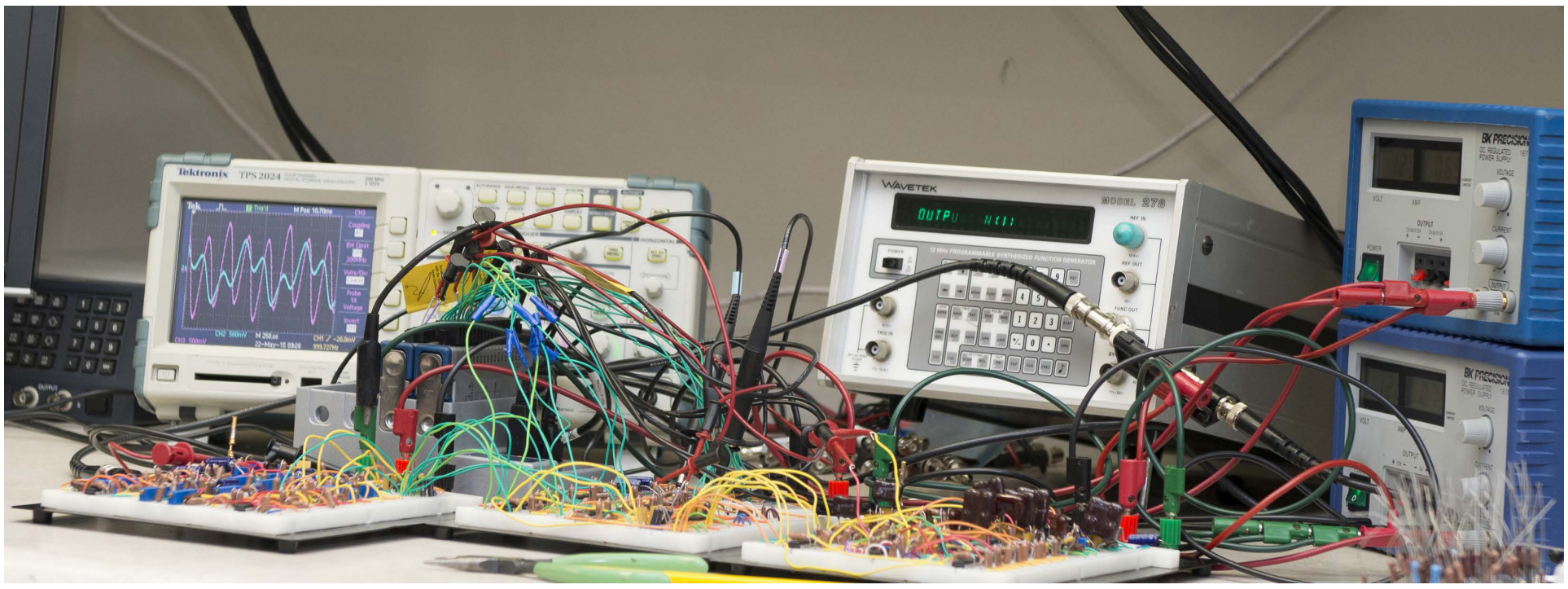

Fig. 1. Photograph of the current hardware setup. The three breadboards correspond to basis signal generation (left), state synthesis (center) and gate operations (right). The devices in the background are an oscilloscope (left), a signal generator (center) and a DC power supply (right). The electronics are interfaced via a desktop computer (to the left, not shown).

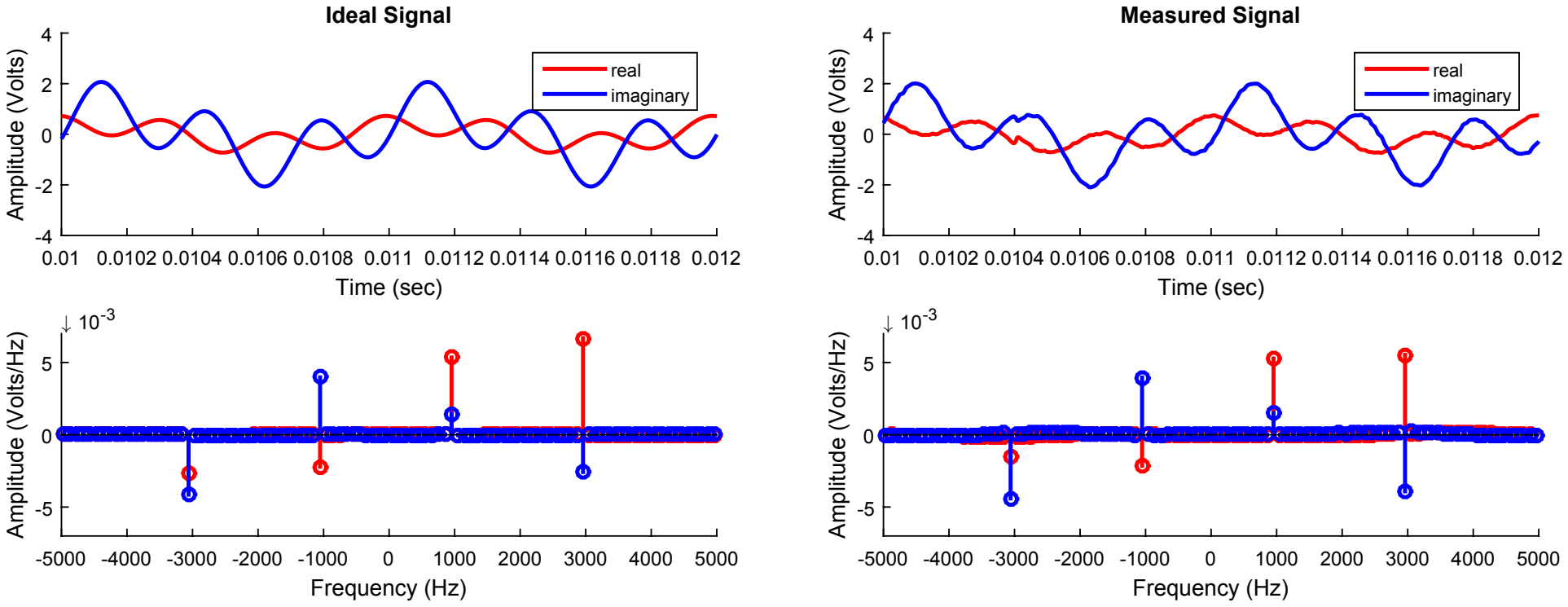

Fig. 2. (Color online) Plot of two complex signals representing a quantum state with coefficients $\alpha_{00} = 0.6579 - 0.2895j$, $\alpha_{01} = 0.5385 + 0.1383j$, $\alpha_{10} = -0.2280 + 0.3953j$ and $\alpha_{11} = -0.2460 - 0.4277j$. The left plot shows the ideal signal, while the right plot shows that recorded signal generated by the hardware. The colors red and blue indicate the real and imaginary parts, respectively. The top plots show the time-domain signals, while the bottom plots show the frequency-domain signals.

State synthesis is performed by multiplying these four basis signals by four complex coefficients $\alpha_{00}$, $\alpha_{01}$, $\alpha_{10}$ and $\alpha_{11}$, each represented by pairs of DC voltages, and adding the results to produce the final, synthesized signal $\psi(t)$ representing the quantum state $|\psi\rangle$. An example of a synthesized signal is given in Fig. 2, which shows the ideal pair of signals, representing the real and imaginary parts of $\psi(t)$, and the recorded signals generated in the hardware. In this example, the state is specified by the complex coefficients $\alpha_{00} = 0.6579 - 0.2895j$, $\alpha_{01} = 0.5385 + 0.1383j$, $\alpha_{10} = -0.2280 + 0.3953j$ and $\alpha_{11} = -0.2460 - 0.4277j$.

To implement gate operations, we use a set of analog four-quadrant multipliers, filters and operational amplifiers to realize the mathematical operations described previously. For example, to perform a gate operation on qubit-$A$ we use a pair of low-pass filters to remove the 2000 Hz component from $e^{\pm j\omega_A t}\psi(t)$. The resulting partial projections $\psi_0^{(A)}(t)$ and $\psi_1^{(A)}(t)$ are a pair of 1000 Hz signals corresponding to qubit-$B$. To perform the gate operation, we take a matrix $U$, given by, say

$$U = \begin{bmatrix} U_{00} & U_{01} \\ U_{10} & U_{11} \end{bmatrix} = \begin{bmatrix} 0.1759 + 0.1836j & 0.4346 + 0.8460j \\ -0.4346 + 0.8640j & 0.1759 - 0.1836j \end{bmatrix} \tag{14}$$

and use it to construct two qubit-$A$ signals of the form $U_{00}e^{j\omega_A t} + U_{10}e^{-j\omega_A t}$ and $U_{01}e^{j\omega_A t} + U_{11}e^{-j\omega_A t}$. These, in turn, are multiplied by the corresponding partial projections and added to form the final signal $\psi'(t)$, given by

$$\psi'(t) = (U_{00}e^{j\omega_A t} + U_{10}e^{-j\omega_A t})\psi_0^{(A)}(t) + (U_{10}e^{j\omega_A t} + U_{11}e^{-j\omega_A t})\psi_1^{(A)}(t). \tag{15}$$

The resulting output using the gate specified in Eq. (14) applied to the signal in Fig. 2 is shown in Fig. 3.

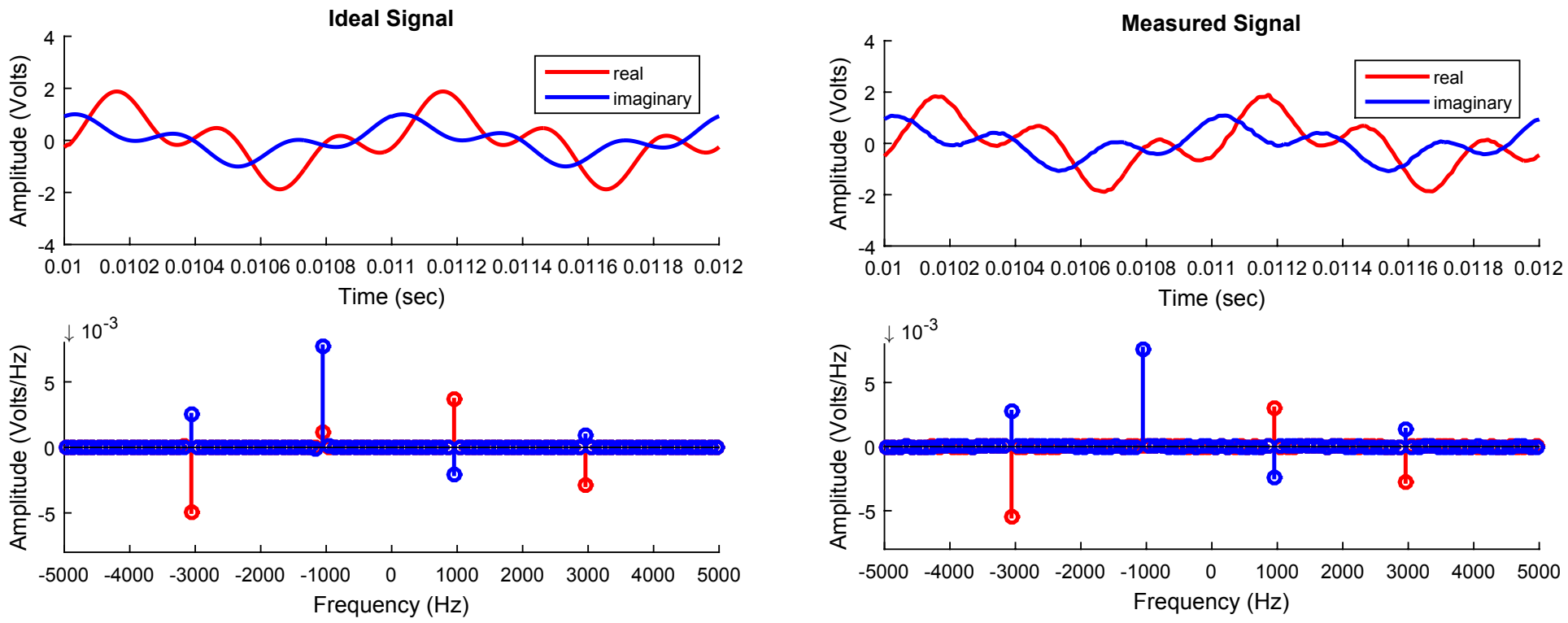

Fig. 3. (Color online) Plot of two complex signals representing the output of a quantum gate operation. The input signal is that shown in Fig. 2, and the gate, operating on qubit-$A$, is given by Eq. (14).

## 4. Fidelity Analysis

The quality of a quantum state or gate operation is typically measured in terms of the *gate fidelity*, which is a number between 0 and 1, where 1 is ideal. For an ideal state $|\psi\rangle$ and recorded state $|\hat{\psi}\rangle$, the fidelity is

$$F(\hat{\psi}, \psi) = \frac{|\langle\hat{\psi}|\psi\rangle|}{\|\hat{\psi}\|\|\psi\|}. \tag{16}$$

Using this definition, we can measure the fidelity of a state synthesis or gate operation over an ensemble of random realizations.

As an illustration, we performed synthesis of the entangled singlet state $|\psi\rangle = [|01\rangle - |10\rangle]/\sqrt{2}$ and examined the fidelity of the signal used to emulate this state (just prior to performing a gate operation on it). Using Eq. (16), we compared the ideal quantum state to the actual signal, using the recorded signal to compute the inner product $\langle\hat{\psi}|\psi\rangle$ and the normalization $\|\hat{\psi}\|$. Figure 4 shows the results of this analysis, where a histogram of fidelity over 500 realizations of the emulated signal is shown. In this example, we find a mean state fidelity of $0.991 \approx 99\%$.

The definition of fidelity given by Eq. (16) assumes a pure initial and final state; in general, the states may be mixed. A mixed state may be thought of as a random ensemble of pure states; for $n$-qubit states a mixed state may be represented by a $2^n \times 2^n$ positive semi-definite matrix $\rho$. In our classical emulation, a single pure state is always realized and can be known, but in a true quantum system this is not so. Instead, one must infer the quantum state through a variety of measurements. One widely accepted approach uses quantum state tomography (QST) to estimate the quantum state from a complete set of orthonormal measurements. In our case, we use the 16 pair-combinations of four Pauli spin matrices, normalized to unity with respect to the Hilbert–Schmidt inner product.[17] To fairly compare our system with a true quantum system, then, we can also perform QST to obtain an estimated quantum state $\hat{\rho}$ and thereby compute the fidelity compared to the ideal quantum state $\rho = |\psi\rangle\langle\psi|$ using the formula[18]

$$F(\hat{\rho}, \rho) = \frac{\mathrm{Tr}(\sqrt{\sqrt{\rho}\,\hat{\rho}\sqrt{\rho}})}{\sqrt{\mathrm{Tr}(\rho)\mathrm{Tr}(\hat{\rho})}}\,. \tag{17}$$

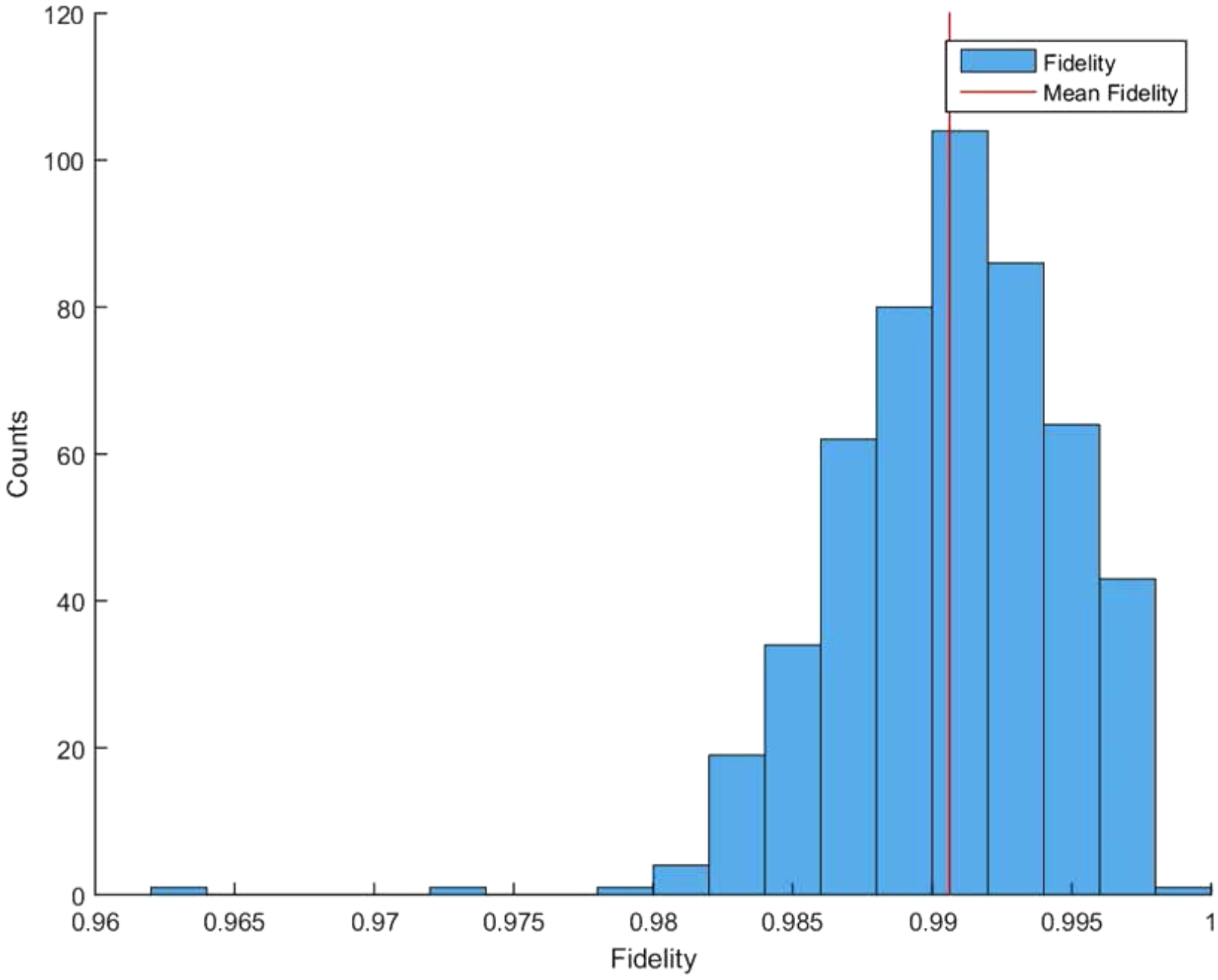

Fig. 4. (Color online) Histogram of signal fidelity for 500 realizations of an initial entangled state of the form $|\psi\rangle = [|01\rangle - |10\rangle]/\sqrt{2}$. The mean fidelity, indicated by the red vertical line, is about 0.991.

To perform such measurements, we take the "bare" pure state and form an ensemble of "dressed" states by rescaling the signal and adding a random noise term. Thus, if $\boldsymbol{\alpha} \in \mathbb{C}^{2^n}$ specifies the bare pure state, then the dressed state is given by $\mathbf{a} = \mathbf{s}\boldsymbol{\alpha} + \boldsymbol{\nu}$, where $s = \sqrt{2} - 1$, $\boldsymbol{\nu} = \mathbf{z}/\|\mathbf{z}\|$, and $\mathbf{z}$ is a standard complex Gaussian. Using these dressed states, a series of measurements are performed using amplitude threshold crossings of projections onto the basis states of the observables in a manner described elsewhere.[16,19] These measurement outcomes are then used in a maximum likelihood QST method to obtain the estimated quantum state $\rho'$ and, from this, the measured fidelity $F(\rho', \rho)$.[20] For the ideal singlet state and a sample size of 1000, for example, we obtained a measured fidelity of about $0.998 \approx 99\%$, comparable to what was found earlier through a direct calculation of $F(\psi', \psi)$.

A similar technique was used to measure gate fidelity. Given a pure singlet state, we applied a random ensemble of unitary gates on qubit-$A$. For each realization of a gate $U$, the ideal quantum state is $|\psi'\rangle = U|\psi\rangle$. If we denote the recorded state by $\hat{\psi}'$, then the gate fidelity will be $F(\hat{\psi}', \psi')$. The results for this example are summarized in

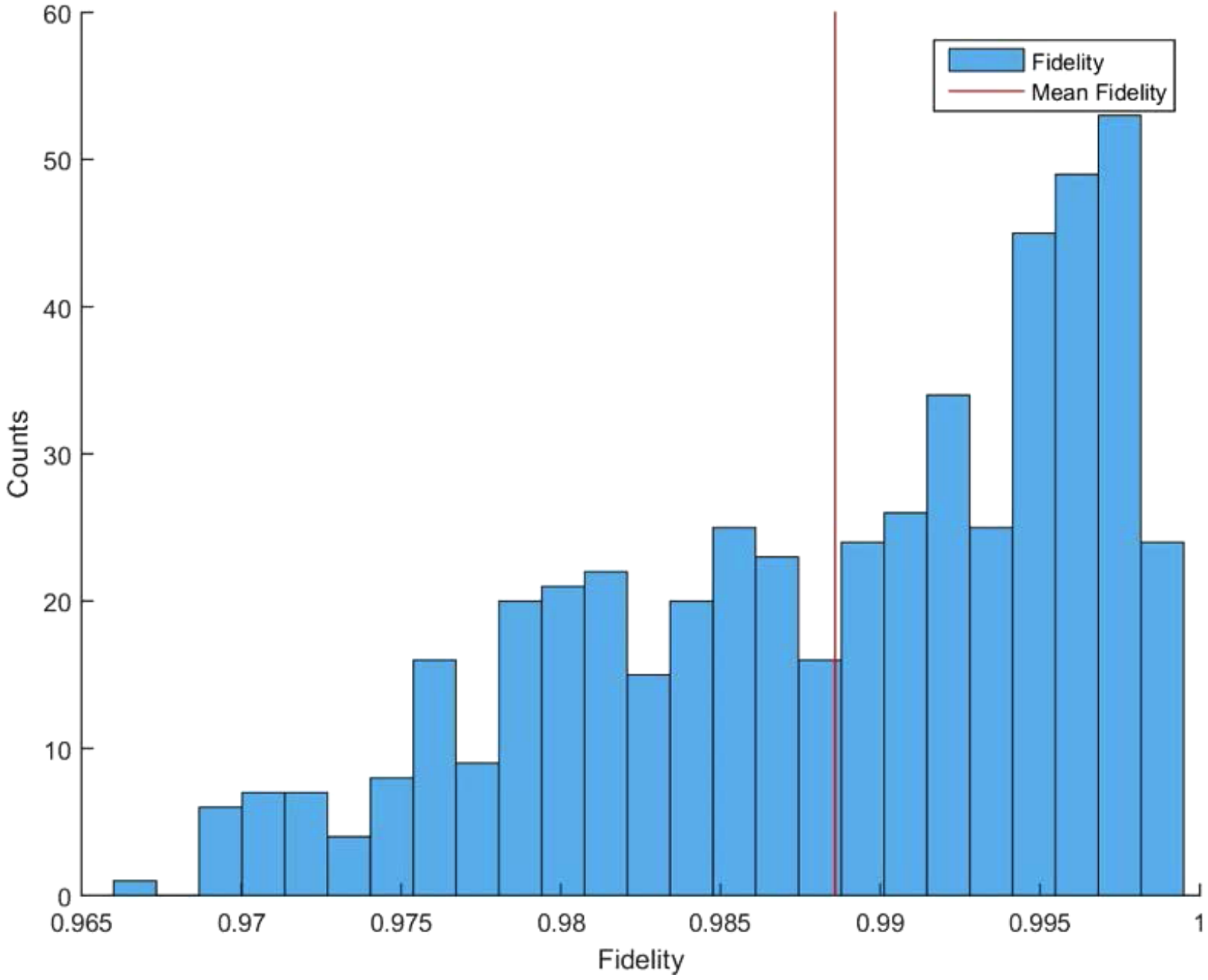

Fig. 5. (Color online) Histogram of gate fidelity for a random set of 500 unitary gates applied to qubit-$A$ on an initial entangled state of the form $|\psi\rangle = [|01\rangle - |10\rangle]/\sqrt{2}$. The mean fidelity, indicated by the red vertical line, is about 0.989.

Fig. 5 where, over an ensemble of 500 runs, a mean fidelity of $0.989 \approx 99\%$ was obtained.

## 5. Practical Considerations

The device as currently implemented is limited to two qubits. Additional qubits are straightforward to implement but require additional bandwidth that scales exponentially with the number of qubits. The complexity of the filters needed to perform the subspace projection operations increases similarly. Using current integrated circuit technology, a device of up to 20 qubits could easily fit on single chip, while 40 qubits would seem to be a practical upper limit.[14]

Current development efforts are focused on building a set of programmable one- and two-qubit gates. Once these are completed, we will have the ability to initialize an emulated quantum state and operate on it with a programmable sequence of universal gates to execute any particular quantum computing algorithm. Future work will focus on scaling up the number of qubits and migrating from a simple breadboard setup to a more sophisticated, chip-based implementation. Classical error sources, such as additive noise, can be modeled as a quantum operations, such as a depolarizing channel, and will be the subject of future investigations.

We envision a device based on the concepts described above that would interface with a traditional digital computer and serve as an analog co-processor, much as is done in our current prototype. Thus, a digital computer, tasked with solving a particular problem, perhaps as a subroutine to a larger computation, would designate an initial quantum state and sequence of gate operations to be performed on this state through a digital-to-analog converter (DAC) interface. The co-processor would produce a final state (i.e. a signal) which would then be subject to a sequence of measurement gate operations. The result would be a particular binary outcome, which would then be reported back to the digital computer via an analog-to-digital (A/D) converter.

The price paid for the computational efficiency of the analog co-processor lies in the hardware complexity needed to implement the device. Each single-qubit projection operation on an $n$-qubit state requires a pair of distinct, comb-like filters with $2^n/4$ (positive) pass-band frequencies, while each two-qubit operation requires the ability to perform $n(n-1)$ different projection operations. A key part of the development for a larger scale device would consist of the design of tunable band-pass filters with multiple nulls for the subspace projection operations.

The number of qubits will also be limited by the available bandwidth and the lowest frequency $\omega_0$ (or period $T$) of the signal. Under the octave spacing scheme, $n$ qubits would require a frequency band from $\omega_0$ to $2^n\omega_0$. Each gate operation would require a time $\mathcal{O}(T)$ to complete, and any useful algorithm would have a number of gates that grows only polynomially in $n$. Thus, for a base qubit frequency of, say, 1 MHz, a single gate operation acting simultaneously on all $2^{10}$ digital states of a 10-qubit signal would take about $T = 1\,\mu$s. If we compare this to a nominal single-core, 1 GHz digital processor, the time to process all $N = 2^{10} \sim 10^6$ inputs would also be about $1\,\mu$s. Thus, a mere 10 qubits would give a processing step-time comparable to that of a modern digital processor.

## 6. Conclusions

The power of quantum computing lies ultimately in the ability to operate coherently on arbitrary superpositions of qubits representing the quantum state. We have shown that the fundamental mathematics of gate-based quantum computing can easily be represented classically, and practically implemented electronically. Thus, in all ways such as device is capable of faithfully emulating a truly quantum system, albeit one of limited scale. This has importance both from a foundational and practical perspective. Foundationally, the work we have described here serves to illustrate that many aspects of quantum computing thought to be both important to its efficacy and uniquely quantum in nature can, in fact, be emulated in an entirely classical manner. Practically, we have shown that by leveraging the concepts of quantum computing and applying them to classical analog systems, one can construct a relatively small-scale device that would actually be competitive with current state-of-the-art digital technology.

## Acknowledgments

This work was supported by an Internal Research and Development grant from Applied Research Laboratories, The University of Texas at Austin and is the subject of a provisional patent. Additional support was provided by the Office of Naval Research under Grant No. N00014-14-1-0323.

## References

1. P. Benioff, *J. Stat. Phys.* **22** (1980) 563.
2. D. Deutsch, *Proc. Roy. Soc. Lond. A* **400**(1818) (1985) 97.
3. P. W. Shor, Algorithms for quantum computation: Discrete logarithms and factoring, in *35th Annual Symp. on Foundations of Computer Science, 1994 Proc.*, November 20–22, 1994 (IEEE, Santa Fe, 1994), pp. 124–134.
4. P. W. Shor, *Phys. Rev. A* **52** (1995) R2493.
5. A. M. Turing, *Proc. Lond. Math. Soc.* **42** (1936) 230.
6. S. Aaronson, *Quantum Computing Since Democritus* (Cambridge University Press, UK, 2013).
7. M. A. Nielsen and I. L. Chuang, *Quantum Computation and Quantum Information* (Cambridge University Press, Cambridge, 2000).
8. M. W. Johnson *et al.*, *Nature* **473** (2011) 194.
9. R. Jozsa and N. Linden, *Proc. R. Soc. Lond. A* **459** (2003) 2011.
10. D. K. Ferry, R. Akis and J. Harris, *Superlattices Microstruct.* **30**(2) (2001) 81.
11. M. Fujishima, K. Saito and K. Hoh, *Jpn. J. Appl. Phys.* **42** (2003) 2182.
12. L. Kish, Quantum computing with analog circuits: Hilbert space computing, in *Smart Electronics, MEMS, BioMEMS, and Nanotechnology*, SPIE, March 3, 2003, Keynote talk.
13. D. Dragoman and M. Dragoman, *Quantum-Classical Analogies* (Springer, Germany, 2004).
14. B. R. La Cour and G. E. Ott, *New J. Phys.* **17** (2015) 053017.
15. B. R. La Cour, *Phys. Rev. A* **79** (2009) 012102.
16. B. R. La Cour, *Found. Phys.* **44**(10) (2014) 1059.
17. G. J. Murphy, *$C^*$-Algebras and Operator Theory* (Academic Press, New York, 1990).
18. R. Jozsa, *J. Mod. Optics* **41** (1994) 2315.
19. B. R. La Cour and E. C. G. Sudarshan, *Phys. Rev. A* **92** (2015) 032302.
20. J. B. Altepeter, E. R. Jeffrey and P. G. Kwiat, *Advances in Atomic, Molecular and Optical Physics*, Vol. 52 (Elsevier, The Netherlands, 2006).